# Interaction of a Hydrogen Refueling Station Network for Heavy-Duty Vehicles and the Power System in Germany for 2050

Philipp Kluschke[1], Fabian Neumann[2]

[1] Fraunhofer Institute for Systems and Innovation Research ISI, Breslauer Strasse 48, 76139 Karlsruhe, Germany
[2] Karlsruhe Institute of Technology, Hermann-von-Helmholtz-Platz 1, 76344 Eggenstein-Leopoldshafen, Germany

*corresponding author: philipp.kluschke@isi.fraunhofer.de, phone: +49 151 5851 9987

## Abstract

A potential solution to reduce greenhouse gas (GHG) emissions in the transport sector is to use alternatively fueled vehicles (AFV). Heavy-duty vehicles (HDV) emit a large share of GHG emissions in the transport sector and are therefore the subject of growing attention from global regulators. Fuel cell and green hydrogen technologies are a promising option to decarbonize HDVs, as their fast refueling and long vehicle ranges are in line with current logistic operation concepts. Moreover, the application of green hydrogen in transport could enable more effective integration of renewable energies (RE) across different energy sectors. This paper explores the interplay between HDV Hydrogen Refueling Stations (HRS) that produce hydrogen locally and the power system by combining an infrastructure location planning model and an energy system optimization model that takes grid expansion options into account. Two scenarios – one sizing refueling stations in symbiosis with the power system and one sizing them independently of it – are assessed regarding their impacts on the total annual energy system costs, regional RE integration and the levelized cost of hydrogen (LCOH). The impacts are calculated based on locational marginal pricing for 2050. Depending on the integration scenario, we find average LCOH of between 5.66 euro/kg and 6.20 euro/kg, for which nodal electricity prices are the main determining factor as well as a strong difference in LCOH between north and south Germany. Adding HDV-HRS incurs power transmission expansion as well as higher power supply costs as the total power demand increases. From a system perspective, investing in HDV-HRS in symbiosis with the power system rather than independently promises cost savings of around one billion euros per annum. We therefore conclude that the co-optimization of multiple energy sectors is important for investment planning and has the potential to exploit synergies.

# 1    Introduction

## 1.1    Motivation

A strong reduction of greenhouse gas (GHG) emissions is required to limit the impact of global warming on humanity and the environment (IPCC, 2013). The transport sector is a major emitter of $CO_2$, accounting for around 23% of the total global energy-related emissions of carbon dioxide in 2014. Within this sector, heavy-duty vehicles (HDV) make up a significant and growing share of approximately 40% (Miller and Facanha, 2014).

A potential solution for reducing GHG emissions in this sector is the use of alternative fueled vehicles (AFV) accompanied by a suitable alternative fuel station (AFS) infrastructure. Developing such infrastructure is a major undertaking that incurs massive investments. At the same time, this is also expected to influence infrastructure requirements in the power system due to increased power supply needs.

Despite a high uncertainty on the right decarbonization technology for HDVs (Kluschke et al., 2019a), most studies say that FCEV powertrains are most suitable for long-haul applications due to comparatively short refueling times and light weight energy storage (Bründlinger et al., 2018; Kasten et al., 2016; Siegemund et al., 2017).

In case of local fuel production, new zero-emission AFS have a potentially strong impact on the local and national energy system. This integration of the transport sector energy demand into the energy sector, called sector coupling, has multiple implications such as grid bottlenecks, changes in the total energy system cost, or regional RE integration.

## 1.2    Literature

Studies of AFS infrastructure modeling have shown that demand-driven location methods outperform strategic location methods on weekly energy transfer (Helmus et al., 2018). The research field of infrastructure cost modeling focuses primarily on the facility location problem from a demand-driven perspective. Here, the flow refueling location problem dominates the research on road transportation that is defined in flow refueling location models (FRLM) (Kuby and Lim, 2005). The FRLM is based on the work of Hodgson (1990), and considers traffic as a demand flow that starts, ends or passes by businesses which seek to serve a given demand. Hodgson suggests using origin - destination (OD) trips to derive the total (refueling) demand flow. From these OD trips, a path along (multiple) nodes is constructed, at which candidate AFS facilities are located, e.g. charging or refueling stations. On a highway network, for example, nodes can be referred to as highway entries, exits or intersections. The FRLM also considers a limited range of vehicles passing along a path (Capar et al., 2013; Kuby and Lim, 2005; Wang and Wang, 2010) und has mainly been applied to passenger car AFS infrastructures (Hosseini and Mir-Hassani, 2017; Jochem et al., 2016; Zhang et al., 2017). Despite the growing interest in AF-HDVs as an alternative to diesel vehicles, research on HDV refueling infrastructure is limited. Only a few researchers have analyzed a potential liquefied natural gas infrastructure for HDVs in the United States (Fan et al., 2017), a potential infrastructure build-up and market diffusion for overhead catenary HDVs in Germany (Wietschel et al., 2017), or determined the cost of catenary infrastructure for the Danish freight vehicle market (Connolly, 2017). To date, there has been no research on modelling a large scale FC-HDV infrastructure. However, a significant difference in passenger car and HDV refueling procedures – such as refueling amount – has already been discussed in previous studies (Elgowainy and Reddi, 2018).

The research focus on sector coupling of transport and energy has only emerged more recently, with most publications appearing from 2017 onwards. Studies of sector coupling originate either from an energy sector perspective (more general) or from a transport sector perspective (more specific).

In terms of general sector coupling, there are two exemplary studies by Brown et al (Brown et al., 2019; Brown et al., 2018) that use open-source energy modeling. In the more recent publication, they investigate the synergies of sector coupling and transmission network expansion while considering cross-sectoral and cross-border system integration. Their main findings are that seasonal variations in demand and the variability of renewables are predominantly balanced by power-to-gas applications and long-term energy storage, while a daily balance is achieved by combining solar power with battery-electric (BEV) passenger cars. The earlier study likewise analyzes the joint decarbonization of multiple sectors, namely electricity, land transport, and space and water heating, and evaluates how these interact. The result is that, in a zero-direct emission system, the electricity and land transport sectors are defossilized first and the system costs are even marginally cheaper than today's system.

Within the sector coupling literature from a transport perspective, six publications stand out that consider either direct (BEV or overhead catenary) or indirect (hydrogen) electricity-using technologies. In 2019, Plötz et al. (2019) were the first to analyze the coupling of electrified (catenary) trucks with the energy sector regarding their energy demand and load curves. Focusing on the catenary technology for road electrification, they added non-flexible energy demand (energy is needed in the moment of usage in the truck) but considered the most efficient technology (compared with battery electric or hydrogen trucks). They found overhead catenary trucks have a limited additional load on the total energy system, but could not analyze any flexibility value due to the technology's limits. Sector coupling literature that focuses on hydrogen comes largely from the researchers at the Research Center Jülich in Germany, who tend to focus on determining the levelized cost of hydrogen (LCOH). Robinius et al. (2017) propose a model linking the power and transport sector for Germany and describe a scheme of potential pathways that consist of power-to-X approaches for sector coupling. The resulting LCOH vary from 3.00 euro/kg to 6.40 euro/kg. A second study (Welder et al., 2018) determines a cost-optimal design and operation of future energy systems for power-to-gas scenarios with a focus on Germany. An optimization model representing the energy system is applied to investigate hydrogen use for passenger mobility and industry. The impact of the energy system on the cost of hydrogen is described and optimal LCOH of around 4.35 euro/kg are determined. Emonts et al. (2019) modeled hydrogen infrastructures and passenger car fleets to determine the implications of sector coupling on the favorable hydrogen transport options and potential LCOHs. They find LCOH of between 6.8 euro/kg and 11.5 euro/kg, depending on the scenario. Another publication on sector coupling with hydrogen in transport was published by the Lawrence Berkeley National Laboratory (Wang et al., 2018). This study focuses on the interactions between fuel cell vehicles (FCEV) and electric power systems by combining multiple models (vehicle driving patterns, hydrogen stations and production, and an electricity grid model). Wang et al. found that controllable H2 production systems can facilitate renewable energy deployment. However, their analysis did not have a national scope nor did it determine the value of possible flexibility when operating and sizing hydrogen refueling stations.

## 1.3    Objective

This work extends the current research by combining an infrastructure model with an energy system model. The results of this approach are compared with existing studies and synthesized to address the following research questions:

- *What are the costs of producing and supplying hydrogen to domestic HDV traffic from a comprehensive energy system perspective?*
- *How do planning and operating a potential HDV-HRS network and designing a future power system interact?*
- *What are the benefits of designing production systems at HRS to be flexible in terms of operation and size?*

To the best of our knowledge, this paper is the first to combine an infrastructure model and an energy system model to analyze the interplay of a potential HDV HRS network with the energy system, as well as the first to apply such an approach in a case study with national scope. In short, this paper adds novelty to the existing literature in the following areas (i) the method of coupling a demand-driven infrastructure modeling and an open-source energy system modeling, (ii) the transport segment of HDV (in contrast to the more commonly assessed passenger cars), (iii) the technology, (fuel cells and green hydrogen), and (iv) the analyzed region (Germany).

The paper is structured as follows: First, we present the approach in Section 2 and describe the case study in Section 3. After presenting the results in Section 4, they are discussed in Section 5. We close with conclusions and suggestions for further research in Section 6.

## 2 Methodology: Combining transport infrastructure and energy system modeling

This paper merges the results of a transportation infrastructure model into an energy system model. Each model is described separately below, before the combination of the models is addressed.

### 2.1 Transport infrastructure modeling

First, we introduce the Node-Capacitated Fuel Refueling Location Model (NC-FRLM), which is based on Capar et al. (2013) and was extended in (Kluschke et al., 2019b). This model determines the optimal (fueling or charging) station locations for alternatively fueled vehicles based on traffic flow and considering node capacity limits; e.g. restrictions on the amount of hydrogen a refueling station can provide. There are seven major assumptions applied in this model:

1. A vehicle drives along a single origin-destination (OD) path that is the shortest path from the center of an origin area to the center of a destination area.
2. The traffic volume on a single OD path is known in advance.
3. A station will only be located at a node of the highway network.
4. The distance traveled is proportional to the fuel consumption.
5. Refueling is only necessary for trips longer than 50 km.
6. The drivers have full knowledge about the location of AFS along the path and refuel efficiently to complete a single trip.
7. The maximum driving range that can be achieved for each single refueling process is similar for each vehicle.
8. Each vehicle starts and ends its trip with the same fuel level, which is sufficient to cover the specific trip range.
9. Nodes and AFS are capacitated.

The first four assumptions suit our case well as trucks mostly drive along highway networks from one specific location to another. With regard to the first assumption, the shortest path from the entrance node to the exit node in the highway network is calculated by applying Dijkstra's algorithm (Dijkstra, 1959) to every OD pair. We assume a vehicle completes a single trip rather than a round trip because this better suits our case study of trucks, which normally receive a delivery order to another location once it reaches its destination (tramp traffic) (Gürsel and Tölke, 2017). The assumptions (3) – (5) are made to increase the effectiveness of AFS deployment. We assume 50km to be a suitable cap to balance removing vehicle flow data from the set and incorporating an increasing likelihood of refueling after 30-60 minutes on the road. The sixth assumption is reasonable nowadays and even more applicable to our case study in 2050 since most trucks are now equipped with a decent navigation system technology. AFV-HDV tend to be confined to a comparatively narrow range of technical specifications, especially for fuel cell HDV, which makes the seventh assumption reasonable. The refueling strategy is implied by (8), where we assume no private AFS at the trip's origin or destination. Due to the previously mentioned tramp traffic nature of truck tour planning, we assume the same fuel level at the beginning and end of a trip. Consequently, the total amount a vehicle refuels is proportional to the total distance of its trip. As subsequent journeys are not considered, applying this assumption also prevents excessive refueling and reflects the energy needed to cover the actual trips made. Assumption (9) describes the node capacity limit extension.

The formulation of the node-capacitated FRLM model is then as follows (cf. (Capar et al., 2013; Kluschke et al., 2019b)):

$$Min \sum_{i \in N} z_i \qquad (1)$$

Subject to:

$$\sum_{i \in K_{j,k}^q} z_i \geq y_q, \forall q \in Q, a_{j,k} \in A_q \qquad (2)$$

$$\sum_{q \in Q} [f | q \cdot y_q] \geq S \qquad (3)$$

$$y_q, z_i \in \{0,1\}, \forall q \in Q, i \in N \qquad (4)$$

$$\sum_{q \in Q} [f_q \cdot y_q \cdot r_{iq} \cdot p \cdot g_{iq} \cdot x_{iq}] \leq c \; z_i, i \in N \qquad (5)$$

$$\sum_{i \in K_{jk}^q} x_{iq} = y_q, \forall q \in Q, a_{j,k} \in A_q \qquad (6)$$

$$\sum_{i \in N} \sum_{q \in Q} x_{iq} = y_q \cdot l_q \qquad (7)$$

$$x_{iq} \leq z_i, i \in N, q \in Q \qquad (8)$$

$$0 \leq x_{iq} \leq 1 \qquad (9)$$

**Nomenclature**

**Sets and indexes**

| | |
|---|---|
| $A_q$ | Set of directional arcs on the shortest path q, sorted from the origin to the destination |
| $K_{j,k}^q$ | Set of all potential AFS sites / nodes that can refuel the directional arc $a_{j,k}$ in $A_q$ |
| N | Set of all nodes that form the highway network, N = {1,…n} |
| Q | Set of all OD pairs |
| i,j,k | Indices of potential facilities at nodes |
| q | Index of OD pairs |

| $a_{j,k}$ | Index of unidirectional arc from node j to node k |
|---|---|

**Parameters**

| c | Capacity at node i |
|---|---|
| $f_q$ | Total vehicle flow per OD trip refueled |
| $g_{iq}$ | Indicator of potential station location |
| $l_q$ | Refueling occasion on path q |
| p | Fuel efficiency |
| $r_{iq}$ | Amount of refueling to reach maximum tank (difference between current fuel level and maximum fuel level) |
| S | Objective percentage of refueled traffic flow[1] |

**Decision variables**

| $xi_q$ | Proportion of vehicles on path q that refuel at node i |
|---|---|
| $y_q$ | Parameter that indicates proportion of vehicles refueled on path q |
| $z_i$ | $z_i = 1$ if an AFS is built at node i. $z_i = 0$ if otherwise |

Equation (1) represents the objective to minimize the number of stations built ($z_i$) at all nodes $i$ in the entire network $N$. Equation (2) is a constraint developed by (Capar et al., 2013) to replace the requirement to calculate initial feasible station combinations in most FRLM models. Constraint (2) assures that if path $q$ is refueled ($y_q$), there should be a minimum of one station that is built ($z_i$) at one of the nodes $i$ that lies in the set of potential station sites $K_{j,k}^q$. Equation (3) is a constraint to ensure that the total amount of flow ($f_q$) in all refueled paths ($q$) is larger than or equal to the pre-defined minimum service coverage. Equation (4) defines the nature of variables, where $z_i$ and $y_q$ are binary variables, $q$ is an element of set $Q$, and $i$ is an element of set $N$.

Constraints (5) – (7) are added to limit the capacity per potential station based on the quantity of consumed energy, e.g. fuel. Constraint (5) states that a station can be built if the total demand served at node $i$ is less than the capacity limit. The total demand that is served at node $i$ on path $q$ is equal to the total flow of trucks ($f_q$) multiplied by their fuel consumption ($p$) and the amount of refueling at node $i$ ($r_i$). $g_{iq}$ is a parameter that works as an indicator of potential station location. It is equal to 1 if node $i$ is a potential station on path q and 0 otherwise. Nonetheless, constraint (5) is a quadratic problem, which is difficult to solve. As our main aim is to determine the minimum number of refueling stations required to meet total demand (100% demand coverage), we avoid this problem by setting the variable $y_q$ as a parameter that is equal to 1 and remove constraint (3) accordingly. $x_{iq}$ is a variable that determines whether vehicles on path $q$ should refuel at node i so that the sum of vehicles refueling at node $i$ do not exceed the capacity limit. Constraint (6) defines that if path $q$ is refueled, all vehicles on path $q$ can refuel at any open stations along the path. Constraint (7) ensures the refueling occasion of vehicles on path $q$, which depends on the total distance of the path. Here, $l_q$ is the number of refueling occasions on the OD path $q$, which is calculated by dividing the total distance of OD trip $q$ by the maximum vehicle distance achieved with a single refueling and is rounded up. Constraint (8) ensures that if a vehicle on path $q$ refuels at node $i$, then a station should be invested in. The final constraint (9) defines that $x_{iq}$ should be between 0 and 1.

---

[1]    In this case, S = 100% (all flows will be refueled at least once per trip).

## 2.2 Energy system modeling

The analysis of the energy system, which will incorporate the hydrogen refueling network in addition to the power system, is performed using the electricity system modeling framework PyPSA (Hörsch et al., 2018). PyPSA is open-source software seeking to bridge the gap between power system analysis software and general energy system modeling tools. It combines a multi-period optimal power flow problem with linearized load flow equations and the capacity expansion of generators, energy storage units, and the transmission network infrastructure in a single investment planning problem.

The objective is to find the long-term least-cost energy system comprising the annuitized infrastructure investments plus the short-term costs from generator dispatch over one year, subject to the following set of linear constraints:

1. The energy demand must be met at each location and each point in time.
2. The generator dispatch of renewable generators (such as wind, solar and run-of-river plants) is constrained by temporally and spatially fluctuating availability time series.
3. The dispatch of storage units (such as battery, pumped-hydro, and hydrogen storage) is constrained by their nominal power rating as well as their charging level.
4. The capacity limits of transmission lines must be complied with.
5. The linearized DC power flow equations implementing Kirchhoff's second law must be observed.
6. The installed capacities of generators and storage units may not exceed their geographical potentials.
7. Specified carbon dioxide emission reduction targets must be met.

PyPSA has models for mixed alternating and direct current networks, HVDC links, dispatchable generators as well as generators with time-varying power availability. Moreover, it allows conversion between different energy carriers (e.g. from power to hydrogen) and accounts for efficiency losses as well as inflow and spillage for hydroelectric power plants. Thereby, it is not only capable of pure power system analysis but also a more comprehensive energy system analysis. In such a cross-sectoral setting, the simultaneous co-optimization of generation, storage and transmission infrastructure is pivotal when accounting for the multitude of trade-offs between the variety of energy technologies. The resulting linear optimization problem forms the input to the commercial solver Gurobi, which yields the total annual system costs. In addition to the optimal values of the primal variables, evaluating the dual variables or shadow prices of primal constraints also delivers valuable information such as nodal prices and an endogenous price for carbon dioxide.

Full details on the software package PyPSA and the complete problem definition are presented in Hörsch et al. (2018).

## 3 Case study: data, scenarios and assumptions

### 3.1 Data on heavy-duty vehicles and hydrogen refueling infrastructure

For the NC-FRLM, we use three types of input to characterize German HDV traffic: data to describe the current highway network, individual HDV vehicle trips to understand traffic flow, and a discrete Hydrogen Refueling Station (HRS) station portfolio.

The German Federal Highway Research Institute (German Federal Highway Research Institute, 2019) regularly provides traffic data on German highways. In this paper, we extract 2,500 traffic surveillance points (hereafter referred to as "nodes") including distances between adjacent nodes. These nodes and their connecting routes represent the complete German highway network of about 13,000 km and 121 highways. We enrich these nodes and routes with data from the most recent HDV road traffic census (2017) and spatial data (geographic coordinates and NUTS3[2] areas). The available HDV data include trailer and tractor trucks (26 to 40 tons). For further spatial analyses, we obtained the distance between each node from BASt using the haversine formula.

Individual vehicle flows are essential information for flow refueling location methods. Therefore, we employ data from (Wemuth et al., 2012), which is one of the most comprehensive surveys on road traffic in Germany. This dataset covers 44,393 individual vehicle trips of about 35,200 vehicles and encompasses both the origin NUTS3 area and the destination NUTS3 area. 4,103 trips are completed by HDVs, i.e. trailer and tractor trucks ranging from 26 to 40 tons, which are congruent to the categories of the previously mentioned road traffic census. For this analysis, we considered the 2,655 HDV trips that commence and finish in Germany in different NUTS3 areas. We excluded traffic flows to or from other countries (external flows), because refueling only part of a flow would be ineffective unless AFS were also available to serve the out-of-country portion of the trip. Therefore, our trip data does not include any flows to or from other countries. Applying the described NC-FRLM to these nodes and OD trips, $K_{j,k}^q$s (set of all potential AFS sites) results in 10,374 sets from all 1,495 OD trips. We assume a market 100% diffusion of FC-HDVs until 2050 in our scenarios and consider the higher heating value for hydrogen.

Further, we define a feasible technology portfolio for HDV-HRS using the Heavy-Duty Vehicle Refueling Cost Models (HDRSAM) (Elgowainy et al., 2017). The portfolio sizes S to XXL in Table 1 are based on the arrival frequency of vehicles and German legislation, as there are capacity limitations for nodes and stations defined in national regula-

---

tions. According to the German Federal Immission Control Act (Bundesimmissionsschutz-Verordnung BImSchV, Annex 1), operators storing less than 30t hydrogen at a HRS may use a "simplified procedure" when applying for permission to build the HRS compared to a separate extended "approval procedure with public participation" when planning to store more than 30t hydrogen. In addition, storing more than 30t hydrogen would require "extended obligations" (BImSchV, Incident Ordinance). Hence, we set 30t of hydrogen as the maximum capacity per node for our stations as shown in Table 1 (more details on the economics can be found in the Appendix in Table 6).

| | Parameter | Unit | XS | S | M | L | XL | XXL |
|---|---|---|---|---|---|---|---|---|
| Technology | Heavy-duty vehicles | [HDV/d] | 15 | 31 | 61 | 123 | 246 | 492 |
| | Passenger cars | [PC/d] | 180 | 372 | 732 | 1,476 | 2,952 | 5,904 |
| | Hydrogen demand | [kg/d] | 938 | 1,875 | 3,750 | 7,500 | 15,000 | 30,000 |
| | Low pressure storage | [kg] | 938 | 1,875 | 3,750 | 7,500 | 15,000 | 30,000 |
| | High pressure storage | [kg] | 113 | 225 | 450 | 900 | 1,800 | 3,600 |
| | Electrolyzer | [MW] | 2 | 5 | 9 | 19 | 37 | 74 |

Table 1: Overview HRS types (XS, S, M, L, XL and XXL) based on HDRSAM tool (Elgowainy et al., 2017) and own assumptions.

The demand time series of hydrogen at individual stations shown in Figure 1 is determined by the product of their local annual demand and a normalized time series representing the share of annual demand consumed in each snapshot. The latter is obtained by expanding the hourly driving patterns of heavy-duty trucks in Germany in a typical week to a full year taking seasonal variations into account (German Federal Highway Research Institute, 2019). Note that due to the lack of more appropriate data, this method assumes perfect correlation between refueling patterns and driving patterns and neglects regional variations; i.e. the normalized demand time series is the same for every location.

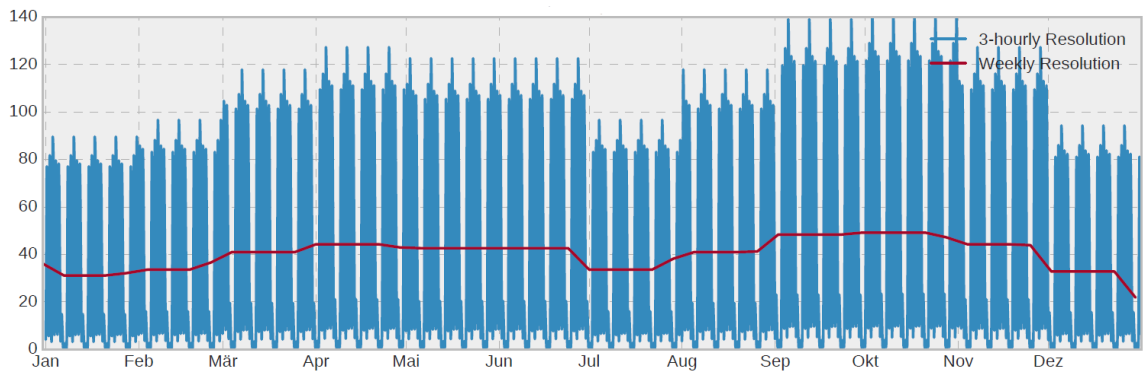

Figure 1: Hydrogen demand per hour of the year (100% demand corresponds to the total annual demand of hydrogen for fuel cell HDVs)

## 3.2 Data on the German power system

Data on Germany's power system infrastructure is taken from PyPSA-Eur, which is an open model dataset of the European power system at the transmission network level (Hörsch et al., 2018). It includes:

- the transmission infrastructure for the ENTSO-E area using the tool Grid-Kit,

- an open database of conventional power plants obtained with the power plant matching tool, which merges multiple publicly available power plant databases,

- spatially and temporally resolved time series for electrical demand derived from a top-down heuristic based on population and gross domestic product,

- spatially and temporally resolved time series for variable renewable generation availability based on weather data for the year 2013 and underlying technical wind turbine and PV module characteristics, and,

- geographic potentials for the expansion of renewable generation based on land eligibility, nature conservation areas and assumptions on allowable densities.

Full details on the routines and data sources of PyPSA-Eur are found in Hörsch et al. (2018).

Since the network of hydrogen refueling stations is limited to Germany in our analysis, we only take an extract of the European model. This results in a network with 333 nodes, in which we disregard electricity imports or exports to adjacent countries and thereby enforce an energy balance within Germany contrary to the current net surplus. The temporal resolution is reduced to two hours for one year yielding 4380 snapshots. This is a compromise between computational tractability on the one hand and considering a large range of operating conditions that are vital to investment planning on the other.

As the outlook towards 2050 exceeds the lifetimes of most existing components, we apply a greenfield planning approach, which largely ignores the current energy system layout and disregards the pathway to the optimal system layout. An exception to this is the AC power transmission infrastructure, for which current electrical characteristics are employed. We further assume no other fossil-fueled generators other than open-cycle (OCGT) and combined-cycle gas turbines (CCGT). Specifically for this case study, we assume that nuclear, lignite and hard coal power plants are phased out under regulatory law by 2050.

The renewable generators considered are solar photovoltaic, run-of-river power plants, and onshore as well as offshore wind farms connected to the mainland by either alternating (AC) or direct (DC) current lines.

In terms of energy storage, hydrogen storage with electrolysis and reconversion in fuel cells and generic batteries is permitted at every node without capacity restrictions. We also consider the pumped-hydro power plants currently in operation.

Transmission lines can be reinforced up to double their current capacity. HVDC link route options are taken from the Ten-Year Network Development Plan (TYNDP) provided by ENTSO-E (ENTSO-E, 2019) and, independently of currently planned capacities, are allowed to expand up to 10 GW in net transfer capacity. To approximate N-1 condition and accommodate for reactive power flows neglected by the linear power flow approximation, we apply a line contingency by setting the maximum line utilization to 70% of its nominal transfer capacity. Table 2 outlines the cost assumptions used for power system assets. For all technologies a universal discount rate of 7% is assumed.

| Asset | FOM [%/a] | VOM [euro/MWh$_{el}$] | Efficiency [%] | fuel [euro/MWh$_{th}$] | lifetime [a] | investment | [unit] |
|---|---|---|---|---|---|---|---|
| HVAC overhead | 2 | | | | 40 | 400 | euro/MW/km |
| HVDC inverter pair | 2 | | | | 40 | 150,000 | euro /MW |
| HVDC overhead | 2 | | | | 40 | 400 | euro/MW/km |
| HVDC submarine | 2 | | | | 40 | 2000 | euro/MW/km |
| CCGT | 2.5 | 4 | 50 | 21.6 | 30 | 800 | euro/kW$_{el}$ |
| OCGT | 3.75 | 3 | 39 | 21.6 | 30 | 400 | euro/kW$_{el}$ |
| Run of river | 2 | | 90 | | 80 | 3000 | euro/kW$_{el}$ |
| Solar PV | 4.17 | 0.01 | | | 25 | 600 | euro/kW$_{el}$ |
| Biomass | 4.53 | | 46.8 | 7 | 30 | 2209 | euro/kW$_{el}$ |
| Onshore wind | 2.45 | 2.3 | | | 30 | 1110 | euro/kW$_{el}$ |
| Offshore wind | 2.30 | 2.7 | | | 30 | 1640 | euro/kW$_{el}$ |

| Pumped hydro storage | 1 | 75 | 80 | 2000 | euro/kW$_{el}$ |
|---|---|---|---|---|---|
| Battery inverter | 3 | 81 | 20 | 323 | euro/kW$_{el}$ |
| Battery storage | | | 15 | 154 | euro/kWh |
| Electrolysis | 4 | 68 | 20 | 510 | euro/kW$_{el}$ |
| Fuel cell | 3 | 58 | 20 | 339 | euro/kW$_{el}$ |
| Hydrogen storage | | | 20 | 19 | euro/kWh |

Table 2: Asset cost assumptions of the power system model

### 3.3    Modeling hydrogen refueling stations in the energy system model

We take the station locations from Section 2.1 and the hydrogen demand time series from Section 3.1 to embed the hydrogen demand of domestic heavy-duty road vehicles and the required geographical distribution of refueling stations into the energy system model of the power sector. We do not presume any capacities of electrolyzers or hydrogen storage units since these will be re-optimized depending on their interaction with the power system. Figure 2 visualizes an overlay of the German power transmission network model and the German highway network.

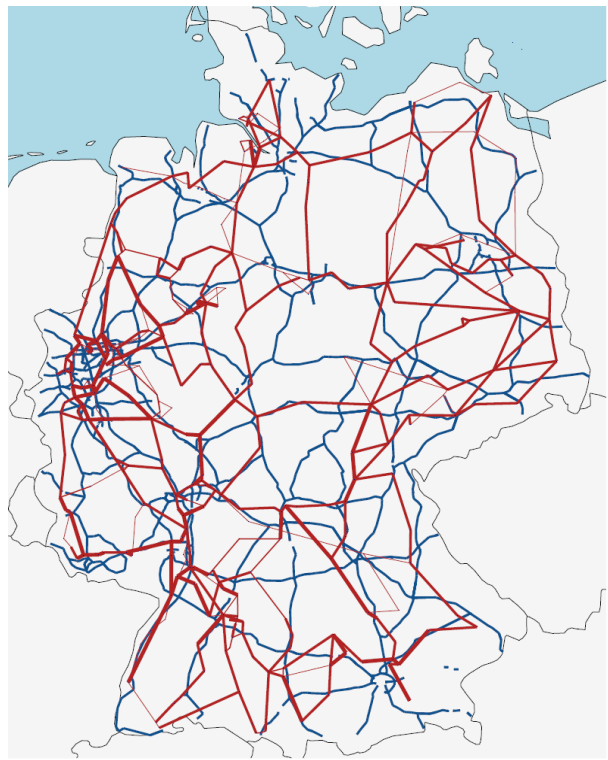

Figure 2: Overlay of German highway network (blue lines) and stylized high-voltage power network (red lines)

In the energy system model, refueling stations are represented by an electrolyzer and a low-pressure hydrogen storage. Our analysis is restricted to on-site electrolysis at the refueling stations; i.e. we do not consider hydrogen transport via pipelines or roads (due to multiple obstacles such as low potential economies of scale of a central hydrogen production having already electrolyzer megawatt scales at the stations, a low likelihood of natural gas pipeline rededication for hydrogen or a complete new hydrogen pipeline network from scratch covering multiple thousand kilometers). Instead, local hydrogen demand must be met by hydrogen produced from the local electrolyzer. Furthermore, while we do allow the reconversion of hydrogen to power in a fuel cell for hydrogen storage options, we do not consider this for hydrogen refueling stations. As stated earlier, for regulatory reasons, the maximum low-pressure hydrogen energy storage capacity is 30 tonnes (at 200 bar), equivalent to 999.9 MWh (cf. Section 3.1). PyPSA-Eur models the power system on the transmission level only and hydrogen refueling stations may not be located in direct proximity to the high-voltage grid. Therefore for reasons of simplicity, we assume the costs of connecting an electrolyzer to the power grid are proportional to its distance to the nearest high-voltage substation measured as a straight line, and use the specific costs for transmission level line types from Table 2.

The capital expenditures for additional components of hydrogen refueling stations that are not linked to the electrolyzer or low-pressure storage capacities and that do not interact with the state of the power system, but are rather a function of total or peak hydrogen demand, are added ex-post; i.e. after the investment planning problem has been solved. These components include compressors (to compress hydrogen from 200 to 1,000 bar), high-pressure hydrogen storage (1,000 bar), dispensers (to transfer hydrogen from high-pressure storage at 1,000 bar toward vehicle storage at 700 bar), safety features and cooling units. Although they are disregarded in investment planning, they constitute a non-negligible cost factor. Cost assumptions for the electrolyzer, low-pressure hydrogen storage and grid connection are identical to those presented in Table 2.

### 3.4    Integration scenarios of hydrogen refueling stations

We investigate the interplay of hydrogen refueling stations with the power system for two different integration scenarios, varying how much hydrogen refueling stations contribute towards cost-efficient power system operation..

In both scenarios the carbon dioxide emissions for the complete modeled system must not exceed 18 Mt/a. This is approximately equivalent to a 95% emissions reduction in the power sector compared to 1990 levels in Germany (German Environment Agency, 2019). It is important to note that additional hydrogen demand from domestic heavy-duty road transport must not incur additional carbon dioxide emissions.

### *Scenario 1: Operational flexibility of HRS*

In this scenario, the operators of hydrogen refueling stations can use how they operate their electrolyzer to minimize the total system costs. This means that the station configuration is not affected by the variable cost of electricity.

Hence, station configuration is determined by locally minimizing the upfront investments in electrolyzers, low-pressure hydrogen storage and grid connection to achieve a feasible operation strategy for onsite electrolysis that can supply the hydrogen demand at each point in time under the given storage restrictions.

### *Scenario 2: Investment and operational flexibility of HRS*

In this scenario, the operators of hydrogen refueling stations can minimize the total system costs by adapting both operation and investments. We assume that the optimal station configuration depends on the varying costs of electricity supply. The hydrogen station operator receives a price signal incentivizing a schedule conducive to cost-efficient power system operation.

Therefore, the investment and operation decisions of the operators of hydrogen refueling stations become part of the global optimization problem to minimize total system costs, which determines the station configuration that incurs the least costs.

# 4 Results

We evaluate the case study results in four consecutive steps. First, we present the optimized system layout of the German power system without hydrogen refueling. This serves as the baseline for comparison with the sector-coupled integration cases. Second, we outline the characteristics of a potential network of hydrogen refueling stations. Third, we discuss the economic parameters of refueling stations and, fourth, we investigate how the optimal power system layout changes if hydrogen refueling stations are incorporated into the power system.

## 4.1 German power system without HDV-HRS

Before adding the HRS network to the power system, we examine what a cost-efficient power system for 2050 could look like without coupling to other sectors, in order to obtain a baseline to which the alternative HRS integration scenarios can be compared. As electrification of other sectors, such as heating, private transport or industry, is not considered in this case study, the total electricity demand is a lower bound. In a fully sector-coupled model, the electricity demand of hydrogen refueling stations would be affected by additional electricity demand resulting from the electrification of other energy sectors. Since all these sectors share Germany's geographical power generation potentials and more electricity is demanded, less favorable sites are chosen for wind and solar farms, and the cost of electricity production will rise.

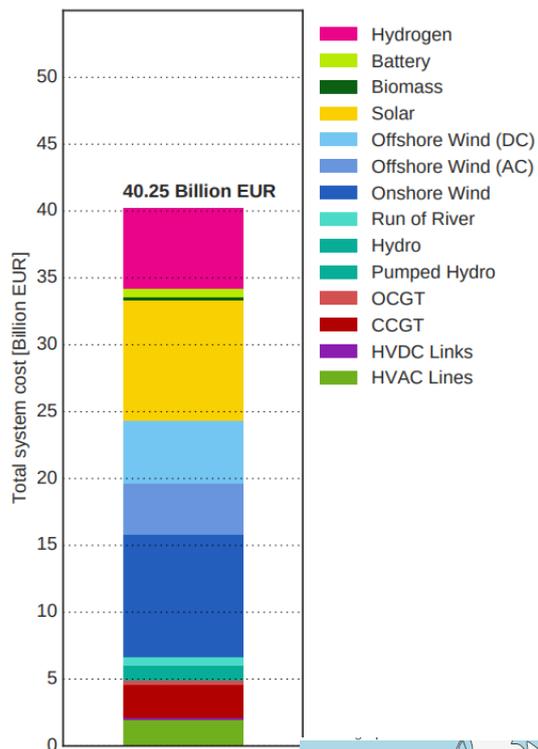

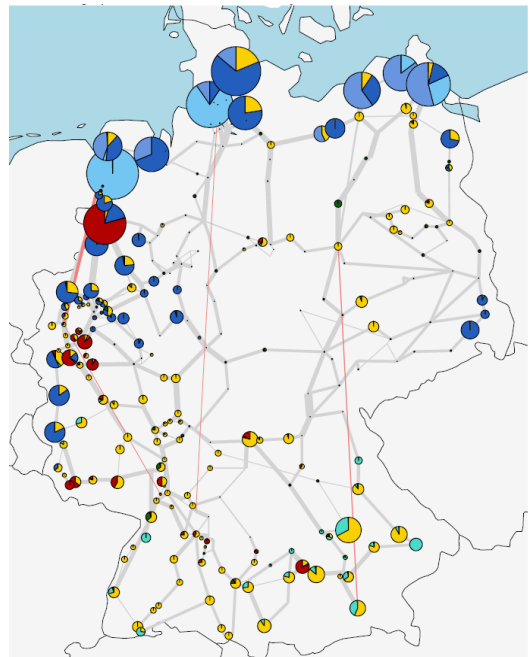

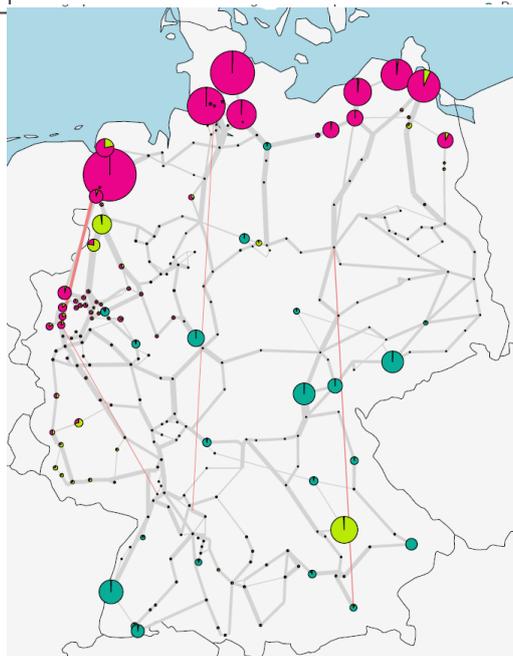

Figure 3: Total system cost of the German power system (left), geographical distribution of electricity production (right) and geographical distribution of storage power capacities (bottom) without HRS network

The total annual system costs amount to 40.25 billion euros per annum, of which about 75% are spent on power generation, 20% on storage infrastructure and 5% on transmission infrastructure as outlined in Figure 3. With a total generation of 518 TWh, the relative total annual system costs amount to 77.6 euro/MWh. Figure 3 further shows that wind turbines dominate power generation, producing 61% of electricity in onshore as well as offshore plants, located predominantly in the north. There is a strong geographical mismatch between generation and load, which is naturally prone to transmission congestion because load centers are mostly based in Southern Germany, , but it is nonetheless the cheapest system layout.

In addition, there is more hydrogen than battery storage deployment. In terms of capital expenditure, the difference amounts to a factor of 10. Hydrogen storage pairs well with locations with high wind power generation, whereas there are only a few but large battery hubs dispersed across the rest of Germany. The most notable is located near Ingolstadt in the South East.

Transmission network expansion is limited to the northern half of Germany and is likewise strongly correlated with wind turbine locations, with lines principally directed south (section Figure 14 in the appendix).

The price that consumers, including hydrogen refueling station operators, pay for electricity is another important feature of the optimization results. We can derive the local cost of electricity production in the system at a particular point in time from the dual variables of the nodal power balance equations that implement Kirchhoff's current law. The value of these dual variables describes the total system costs' sensitivity to consuming an additional unit of power at one location and at one point in time. This value corresponds to the locational marginal prices (LMP) in an idealized market that implements nodal pricing and is capable of factoring in transmission congestion. The increasing deployment of renewables places more strain on the transmission network. This circumstance suggests that grid bottlenecks should be taken into account in electricity markets. . If congestion occurs, nodal prices will vary in the network, but if there were no transmission limits, the nodal prices would be identical at every location. However, the current market structures in Germany with a single bidding zone do not consider internal transmission congestion in the bidding process. Instead, to ensure that the physical limits of transmission are not exceeded, network operators must re-dispatch power stations and curtail renewables to restore order. For the future it is conceivable that re-dispatch is handled through a nodal market approximating LMP. Since we are interested in total system cost effects, we base our analysis on the idealized market design, where consumers pay the locational marginal price, reflecting its impact on total system costs, and leave levies, taxes or other surcharges outside.

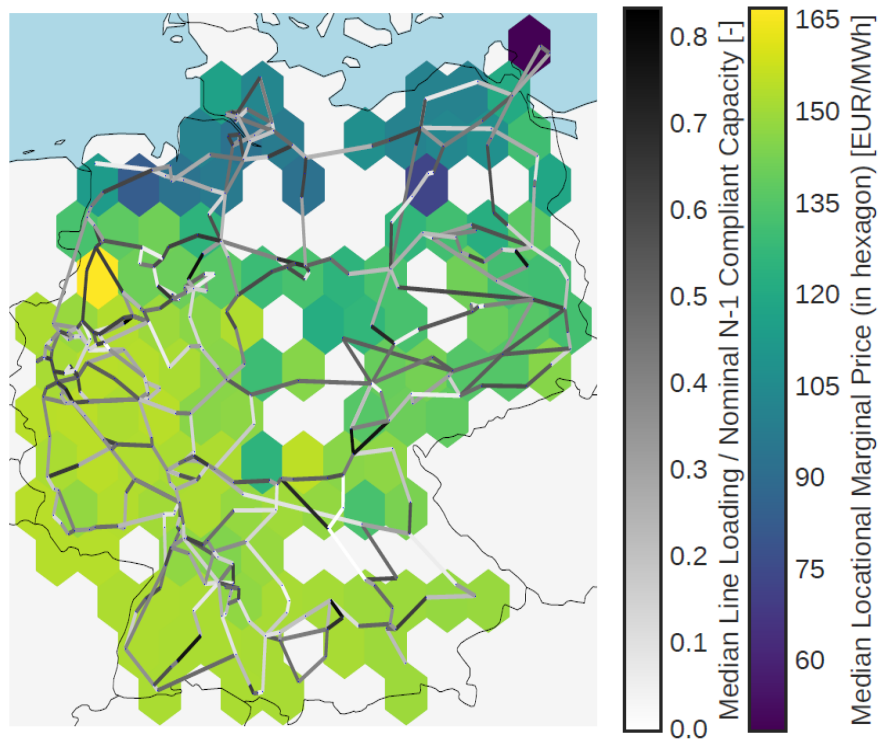

Figure 4: Mean networking loading (left) and local marginal prices of electricity (right)

Figure 4 depicts the median locational marginal prices in combination with average transmission line loadings. We observe a clear North-South differential in median nodal prices between 60 and 165 euro/MWh. This suggests transmission congestion occurs despite capacity expansion and obstructs the flow of low-cost wind power to the South and causes nodal prices to rise, but it has not proven economical to increase the transfer capacities.

Chapter 4.2 evaluates the potential HDV-HRS network providing hydrogen for fuel cell HDVs separately, before chapter 4.3 incorporates the HRS network into the power system.

## 4.2      Potential HDV-HRS network

Presently, there are about 360 conventional gasoline refueling stations located along the German highway network. These stations serve different types of vehicles (passenger cars, light- and heavy-duty vehicles) and are more concentrated in some areas, such as the metropolitan areas around Frankfurt and in Bavaria.

The result of the capacitated FRLM with a supply capacity limit of 30 tons per day is an optimum of 142 stations to serve all the vehicles in all the OD trips shown in Figure 5. Of these 142 stations, 128 stations reach the maximum capacity of 30 tons per day, and the average capacity of all stations built is around 29 tons. The lowest capacity of a station is around 6 tons, which is located on highway A40 around Mülheim an der Ruhr. Around 85% of the stations are located in Western and Southern Germany, which is a result of the high traffic flow and number of OD trips starting and traveling in these regions. We see almost no HRS in Eastern Germany, most probably due to the low industry density and the fact that we did not consider transit road traffic between Germany and neighboring countries in this data set.

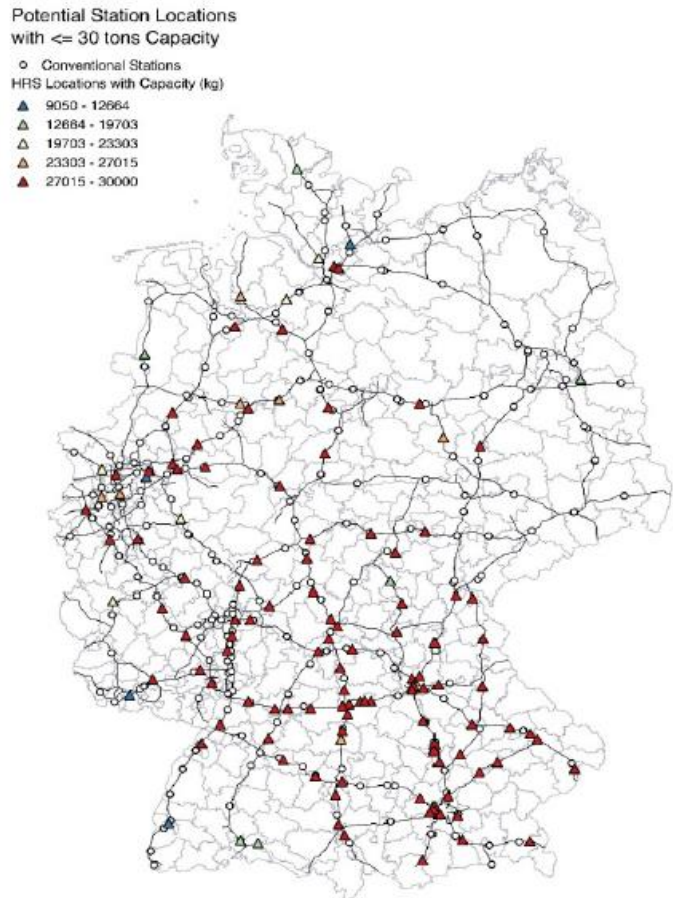

Figure 5: Regional distribution of the existing 360 fuel stations (white circles) and potential 142 HRS locations (triangles) based on the capacitated FRLM with 30t limit

## 4.3 German Power System with HDV-HRS

Having evaluated the power system model and the potential HDV-HRS network separately, we now assess the interplay between the two. First, we evaluate the results from the perspective of the hydrogen refueling stations as we compare the costs of producing hydrogen across the German energy system. Second, we look at how the two integration scenarios shape the cost-optimal system layout and affect total system costs.

### 4.3.1 Implications for HRS network

To appraise and compare the costs of producing hydrogen we use a variant of the Levelized Cost of Hydrogen (LCOH) metric. Conceptionally, the LCOH is very similar to the Levelized Cost of Electricity (LCOE). The LCOH determines the full life-cycle costs of hydrogen production and expresses them as costs per unit of hydrogen produced. The LCOH consists of annuitized capital expenditures (CAPEX) on the one hand, and operational expenditures throughout a year (OPEX) on the other. The CAPEX comprise the initial investments plus fixed maintenance expenditures for the electrolyzers, storage units, compressors and the remaining hydrogen refueling station components, which are converted into annual payments using the components' lifetimes and discount rates. The OPEX comprise the aggregated cost of electricity consumption throughout the year based on the locational marginal prices as well as the variable operation and maintenance costs. Notably, for operating the on-site electrolyzers, we do not reference the LCOE of dedicated generation infrastructure but rather the nodal price of consuming electricity from the power grid, which allows us to express the LCOH related to the total annual system costs. The LCOH is the annual cost of hydrogen production divided by total hydrogen generation, which can be calculated at station level and aggregated or averaged using the annual hydrogen production as a weight.

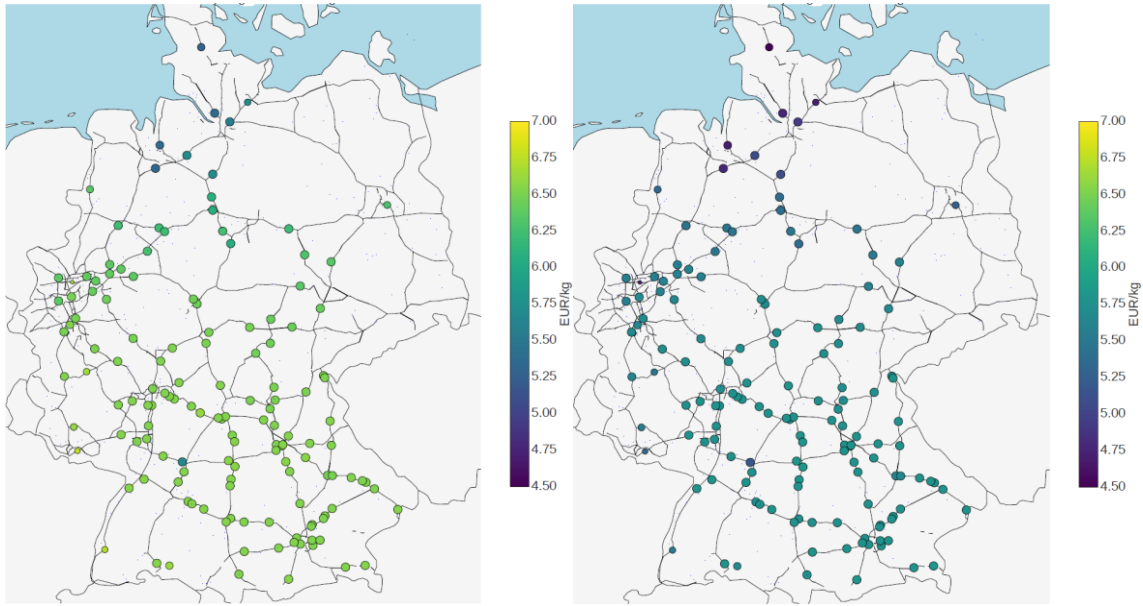

Figure 6: LCOH for scenario 1 (left) and scenario 2 (right)

Sizing hydrogen refueling stations from a system perspective can reduce the average cost of hydrogen production from 6.43 euro/kg to 5.66 euro/kg. This corresponds to a significant reduction by 12%. Figure 6 portrays the geographical distribution of the mean LCOH for each integration scenario. In both scenarios, hydrogen is more expensive in the South than in the North of Germany because of its close links to the average local cost of electricity production as presented in Figure 7. There is a noticeable LCOH deviation near

Heilbronn in the southwest of Germany. This region is located at the end of the HVDC SuedLink route from the north of Germany and can therefore obtain cheap electricity from the north at this node due to the absence of transmission congestion. However, local grid restrictions towards neighboring nodes can substantially increase the locational marginal price of electricity in the node's proximity.

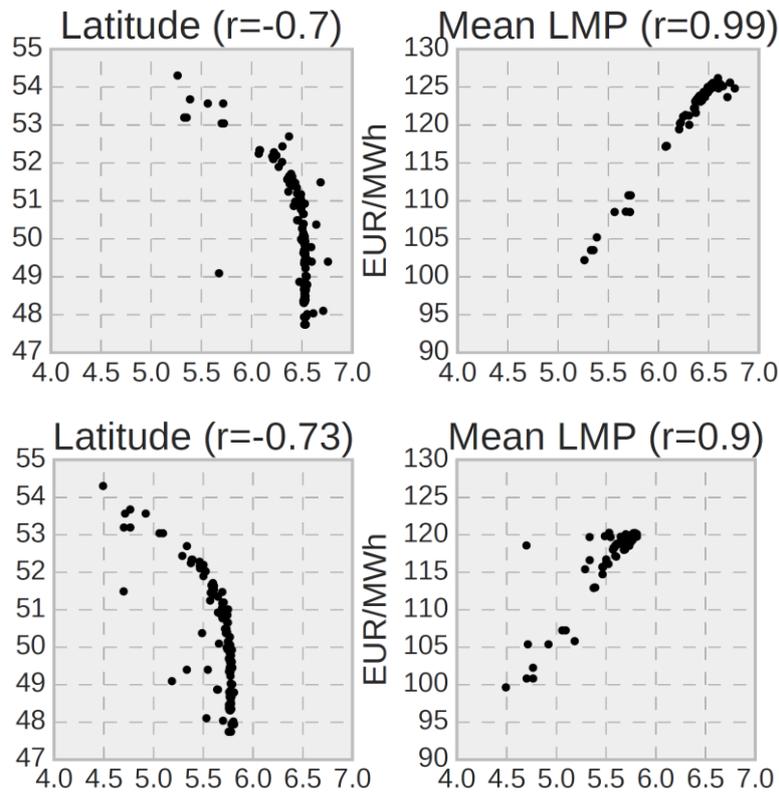

Figure 7: Correlation of LCOH and average cost of electricity production (left) as well as correlation of LCOH and latitude (right) for both scenario 1 (top) and scenario 2 (bottom)

The LCOH is strongly positively correlated with the mean and the variance of the locational marginal price (r=0.99 and r=0.9 respectively with investment flexibility), which, in turn, is closely correlated to the station's latitude. Notably, the difference in LCOH decreases as one moves south. Electrolyzer capacity and total annual hydrogen demand show a lower, but still positive correlation.

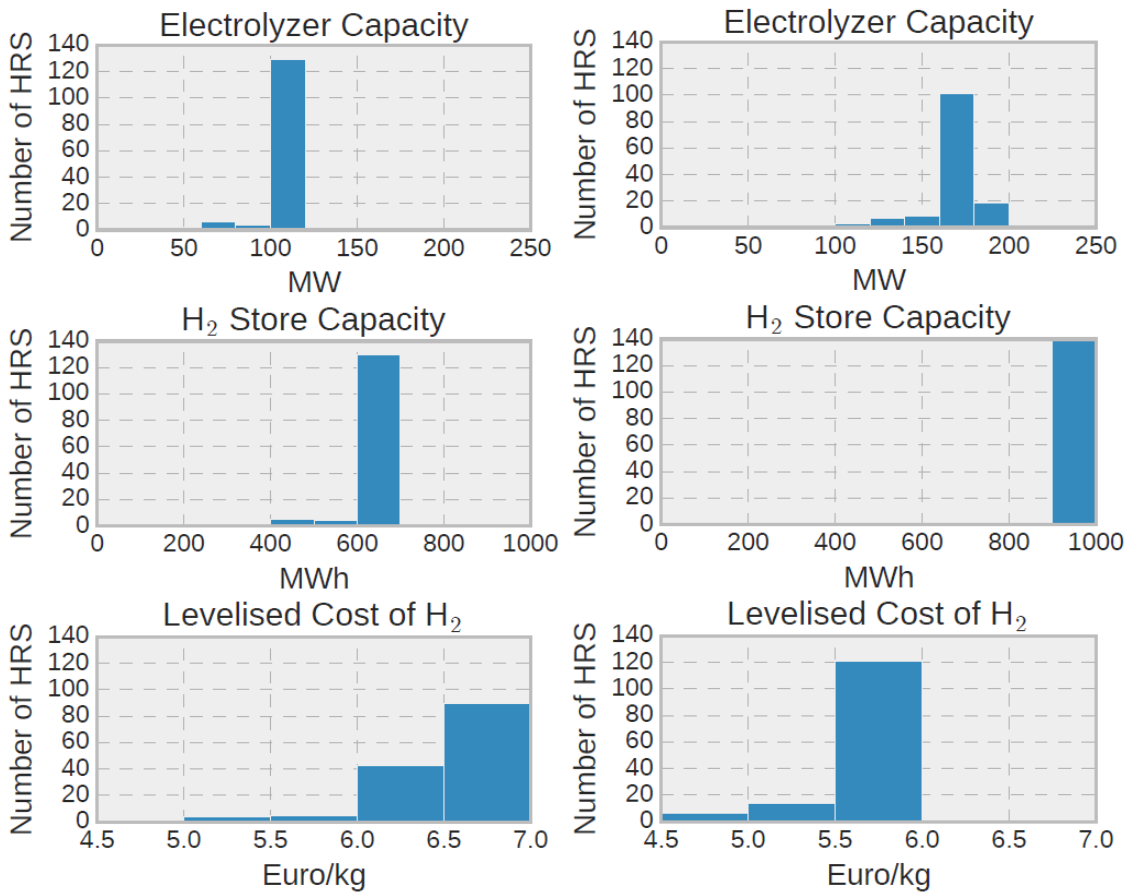

Figure 8: Histograms of electrolyzer capacity (top), hydrogen low pressure storage capacity (middle) and LCOH (bottom) for both scenario 1 (left) and scenario 2 (right)

From the histograms in Figure 8, among other things, we discern a rise in electrolyzer capacities by 69% on average (165 MW instead of 98 MW) in scenario 2 considering investment flexibility. In consequence, the capacity factors of the electrolyzers decline from 0.59 on average to 0.35, because the demand and therefore the amount of hydrogen produced remains constant even though increasing capacities enhance their operational flexibility. Therefore, we may conclude that large electrolyzers are not necessarily a pivotal feature of economical hydrogen refueling stations.

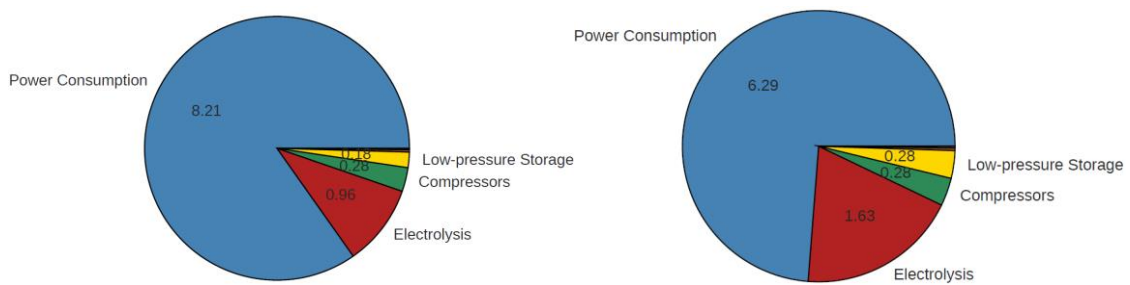

Figure 9: Annualized cost shares of hydrogen refueling infrastructure for scenario 1
(9.69 bn euros p.a. (left)) and scenario 2 (8.53 bn euros p.a. (right))

This change in station configuration is also reflected in Figure 9, which illustrates the share of the individual cost components. With investment flexibility, the capital expenditures for the electrolyzer and low-pressure storage increase by 770 million euros per annum. This raises the CAPEX share in the total cost of hydrogen refueling infrastructure from 15.4% to 26.7%, which is compensated by the significant reduction in power consumption costs of 1.92 billion euros per annum. This highlights that the cost of electricity generation is the main determining factor of hydrogen production costs and that leveraging periods of cheap electricity supply at the expense of oversizing onsite electrolyzers is a sensible economic decision in the assumed market model.

### 4.3.2    Implications for the power system

In both integration scenarios, adding hydrogen refueling stations causes higher total annual system costs in both absolute and relative terms due to the increase in electricity demand of around 72 TWh. The relative costs of electricity are also more expensive. This is due to the fact that the most productive sites for renewable energy generation have already been exploited for decarbonizing the power sector, and additional demand has to be covered by generation at less favorable locations..

The figures referenced in this section are summarized in Table 3 and Table 4.

| | | Without HRS | Scenario 1 | Scenario 2 | Unit |
|---|---|---|---|---|---|
| Demand | Total annual demand | 463.34 | 535.18 | 535.18 | TWh$_{el}$ |
| | Annual electricity demand | 463.34 | 463.34 | 463.34 | TWh$_{el}$ |
| | Hydrogen refueling demand | 0.00 | 71.84 | 71.84 | TWh$_{el}$ |
| HRS | HRS electrolyzers capacity factors | 0.00 | 59.00 | 35.00 | % |
| | LCOH | 0.00 | 6.43 | 5.66 | euro/kg$_{H2}$ |
| | *- CAPEX share* | *0.00* | *15.00* | *27.00* | *%* |
| | HRS electrolyzers | 0.00 | 13.89 | 23.49 | GW |
| | HRS hydrogen storage | 0.00 | 93.02 | 141.99 | GWh |
| Power system | Total annual system costs (relative) | 86.86 | 91.41 | 89.59 | euro/MWh |
| | Total annual system costs (absolute) | 40.91 | 48.92 | 47.95 | bn euro/a |
| | *- HRS Electrolyzers* | *0.00* | *0.96* | *1.63* | *bn euro/a* |
| | *- HRS storage* | *0.00* | *0.18* | *0.28* | *bn euro/a* |
| | *- Other storage (battery, hydro & hydrogen)* | *7.73* | *8.81* | *8.07* | *bn euro/a* |

| | | | | |
|---|---|---|---|---|
| Volume of transmission network expansion (relative) | 17.9 | 22.2 | 21.8 | % |
| Volume of transmission network expansion (absolute) | 7.78 | 9.69 | 9.50 | Twkm |

Table 3: Summary of annual demand, HRS parameters and power system parameters for the German power system without HRS, for scenario 1 and for scenario 2

| | Without HRS | | | Scenario 1 | | | Scenario 2 | | |
|---|---|---|---|---|---|---|---|---|---|
| | Capacity [GW] | Energy [TWh] | Energy [%] | Capacity [GW] | Energy [TWh] | Energy [%] | Capacity [GW] | Energy [TWh] | Energy [%] |
| CCGT | 16 | 45 | 8.6 | 18 | 48 | 7.9 | 17 | 46 | 7.8 |
| OCGT | 6 | 3 | 0.5 | 1 | 0 | 0.1 | 3 | 2 | 0.3 |
| Biomass | 1 | 6 | 1.1 | 1 | 6 | 0.9 | 1 | 6 | 0.9 |
| Offshore wind (AC-connected) | 19 | 72 | 14 | 20 | 74 | 12.3 | 20 | 75 | 12.6 |
| Offshore wind (DC-connected) | 20 | 86 | 16.5 | 20 | 84 | 14 | 20 | 85 | 14.3 |
| Onshore wind | 68 | 157 | 30.2 | 97 | 203 | 33.8 | 85 | 186 | 31.3 |
| Run-of-river | 3 | 18 | 3.5 | 3 | 18 | 3 | 3 | 18 | 3.1 |
| Solar | 163 | 133 | 25.7 | 203 | 167 | 27.9 | 213 | 176 | 29.7 |
| **Total** | **296** | **520** | **100** | **363** | **600** | **100** | **362** | **594** | **100** |

Table 4: Overview of capacities and energy demand per source of renewable energy for the German power system without HRS, for scenario 1 and for scenario 2

Allowing investment flexibility (scenario 2) in addition to operational flexibility (scenario 1) when dimensioning hydrogen refueling stations adds 2 GW of gas turbines and 10 GW of photovoltaics. On the contrary, wind capacity is reduced by 12 GW and hydrogen storage at locations other than at HRS is correspondingly reduced by almost 4 GW. In both cases, batteries play only a negligible role with power capacities below 1 GW. However, when considering the total amount of generation capacities of more than 350 GW, these are still relatively small changes as Figure 10 highlights in terms of the total annual system costs.

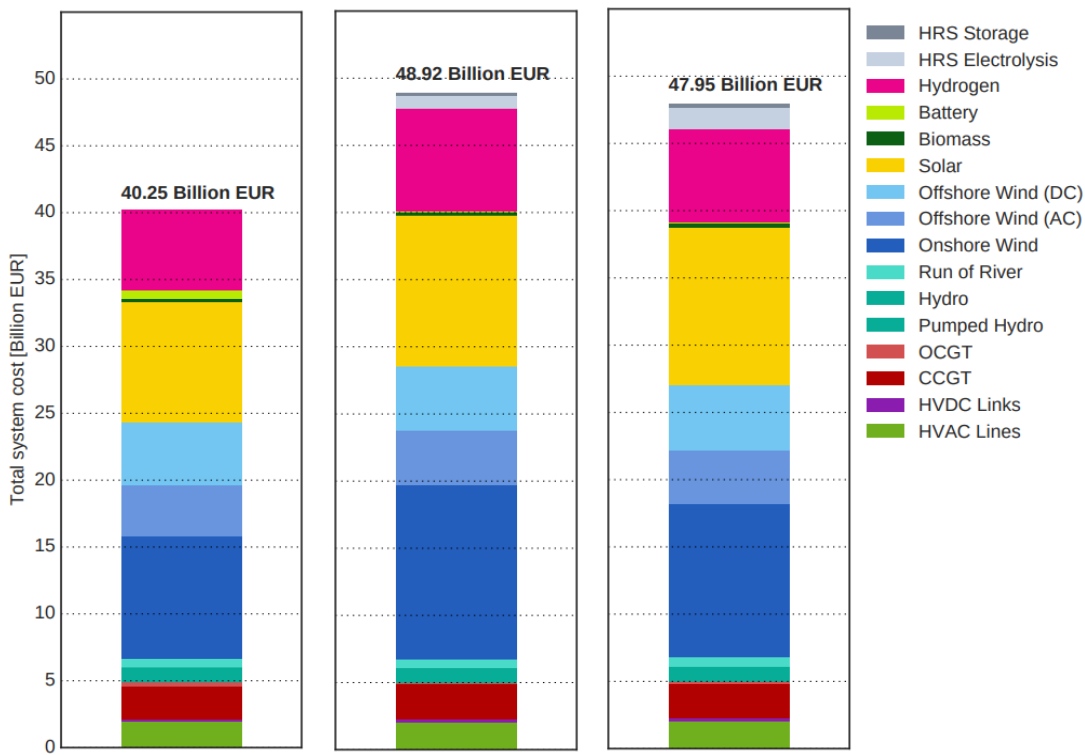

Figure 10: Annual system costs of scenario without HRS (left), scenario 1 (middle) and scenario 2 (right)

A more system-aware planning of HRS can save 2% of total system costs, reducing these from 48.91 billion euro/a to 47.95 billion euro/a. In relative terms, this corresponds to a reduction from 81.52 euro/MWh to 80.80 euro/MWh.

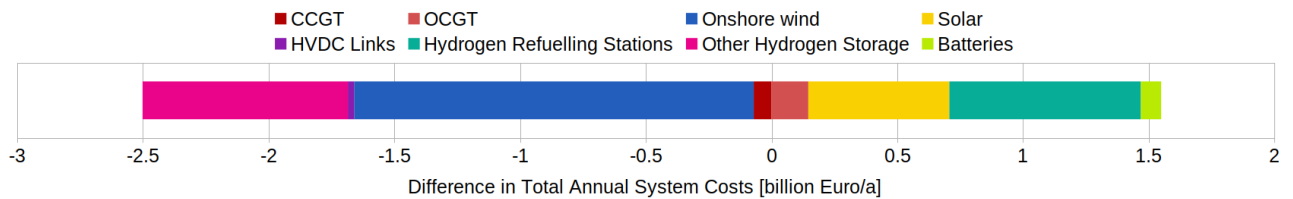

Figure 11: Difference in total annual system costs (scenario 2 compared to scenario 1)

As shown in Figure 11, especially the cost of other storage options can be mitigated by investment planning that takes the entire system into account. The capital expenditures for other storage options drop by 735 million euro/a, while spending on HRS electrolyzers and low-pressure storage amounts to 761 million euro/a needed. This has an overall net positive effect on the total annual system costs because the HRS storage units, which are needed in both cases, can be operated more flexibly in scenario 2.

Table 3 also outlines transmission network expansion in both absolute and relative terms compared to today's transmission system volume. Even without HRS, we have observed considerable line expansion of around 17.9% of today's volume. Adding HRS increases the necessary grid expansion to more than 21.8%, but the differences between the two integration scenarios are rather small (less than 2% difference in absolute volume). As a result, the final network loading levels are only marginally different across the considered scenarios, as shown in the Appendix in Figure 12 and Figure 14. However, adding HRS requires additional network expansion to maintain such loading levels and avoid overexciting network capacities.

# 5 Discussion

We are the first to combine an alternative fuel (hydrogen) network and power system modeling and apply such an approach to real-world data. Our approach allows us to compare the hydrogen costs (LCOH) and electrolyzer operation with current studies from the literature. We reflect more generally on the competitiveness of hydrogen with diesel (regardless of its environmental benefits). Finally, we discuss the observed impact of HRS on the German power system.

Depending on the scenario, our average LCOH vary between 5.66 euro/kg and 6.43 euro/kg with an OPEX share of 70 to 85%. The lowest LCOH are in northern Germany and below 5 euro/kg$_{H2}$ (due to cheaper off-shore wind electricity and no transmission congestion). In comparison, the LCOH of other studies range from 4.40 euro/ kg (note: without HRS cost, Welder et al., 2018), 5.6 euro/kg (middle case, Robinius et al., 2017) to 6.8 euro/kg (Emonts et al., 2019) such that the total LCOH in this study are roughly comparable with existing publications focusing on the integration of fuel cell (passenger) vehicles in the German energy system. However, the composition of LCOH is different compared to other publications, our LCOH contain a lower share of OPEX at below 60%, which indicates other studies must have higher capital costs and lower electricity costs given the similar total LCOH. Notably, other studies dedicated the most attractive renewable capacities to hydrogen production, while our analysis sets renewables for hydrogen production in the context of concurrently decarbonizing the total electricity sector. Considering the different transport focus (heavy-duty vehicles vs. passenger vehicles) of the above mentioned studies, we conclude that total electrification of the transport sector using hydrogen would require significantly more hydrogen to be produced from electricity (ca. 70 TWh for national heavy-duty traffic and ca. 150 TWh for passenger cars (Robinius, 2015) (Seydel, È). Accordingly, LCOH could either benefit from CAPEX economies of scale (although CAPEX only account for 20-30% of total LCOH) or they would become more expensive due to the higher marginal cost of electricity from renewables in Germany, which leads to higher OPEX (which account for 70% to 80% of total LCOH). It is therefore likely that the LCOH we determined are a lower range estimate taking into account the electrification of other transport modes and other sectors as well as the dominance of power consumption costs over initial investments.

When examining hydrogen production (electrolyzer) and distribution, we point out that the current literature only considers centralized hydrogen production, mainly close to the northern coast, and its subsequent distribution via pipelines or trucks, while our analysis focusses on decentralized (on-site) production. When analyzing the potential fuel cell scenarios for passenger cars, the local hydrogen demand seems lower (given 10,000 HRS and 150 TWh cf. Robinius (2015)) than for heavy-duty vehicles (140 HRS and 70 TWh, see section 4.3). In turn, there are lower economies of scale for on-site production of hydrogen at passenger car HRS than for HDV-HRS and on-site production therefore seems more reasonable for HDV-HRS than for passenger car HRS. Notably, other external factors could lead to either an increasing demand for domestically produced hydrogen (e.g. using green hydrogen for the chemical industry) or a decreasing demand (e.g. hydrogen imports from the MENA region).

Addressing the competitiveness of fuel cell HDVs with current diesel HDVs regarding fuel prices, we furthermore compared current diesel prices to estimate competitive hydrogen prices on HDV-HRS (see appendix Table 5). In our (optimal) scenario 2, the resulting maximum hydrogen price of 6.60 euro/kg would allow for 20% of additional cost on top of LCOH (6.60euro/kg – 5.66euro/kg = 0.94euro/kg) for taxes and to operating the HRS including profit margins speculatively assuming that onsite hydrogen production is exempt from network charges. In comparison, the current taxes on diesel account for about 49% and station operations and margins are about 11% (Petroleum Industry Association e.V., 2019). Notably, current diesel prices do not cover the external costs of diesel (environmental pollution due to greenhouse gas emissions and other more locally relevant pollutants) and will therefore most likely increase in the long term, e.g. if the currently discussed $CO_2$ tax is applied in Germany. Further, hydrogen may benefit from tax advantages over diesel. Eventually, however, even with a transition from diesel towards fuel cell HDVs, we perceive the threat of increasing fuel prices for consumers in the future compared with current fuel prices.

Even when considering the most flexible scenario 2, there are significant implications for the power system of an HDV HRS network. The total installed generation capacity increases by ca. 60 GW (equaling +22%), while the total electricity demand increases by about 70 TWh (+13%). However, the relative increase of the system cost (euro/MWh) is only slight by about 3%.

# 6 Limitations

Throughout this paper, we made a series of necessary simplifying assumptions. Based on these limitations, we determine realistic future hydrogen refueling station networks and their role in the energy system.

To determine a suitable HRS network, it is highly desirable to have better data on the driving and refueling profiles of heavy-duty trucks at national level. This involves decoupling driving patterns from consumption patterns as well as retrieving information from a regionally more disaggregated traffic census. Moreover, since German highways accommodate substantial cross-border traffic flows, incorporating these or even extending research to European heavy-duty road traffic would be worth undertaking. Moreover, it should be investigated how our results would change if the restriction of on-site electrolysis were relaxed and transporting hydrogen, e.g. in pipelines, were permitted.

When analyzing the impact on the power grid, it could be investigated how allowing hydrogen reconversion directly at the refueling stations influences the optimal future energy system layout. Furthermore, restricting the analysis to the power sector plus a subset of the transport sector (heavy-duty vehicle freight transport) should be replaced by a sector-coupled model that includes the power, transport, heat and industry sectors, and in which renewable energy potentials are shared across all sectors.

When modelling a suitable network of hydrogen refueling stations, the energy system model should be applied with its natural European scope as the current national scope may underestimate transmission network expansion and overestimate storage expansion, because the benefits of the continental smoothing of renewable feed-in are not covered due to the geographical restriction to Germany and no power exchange between neighboring countries is possible. Moreover, future work should include all the technologies for heavy-duty road transport. Our case study focuses solely on hydrogen and fuel cell technology, although battery-electric or overhead catenary HDVs may also prove suitable technologies for decarbonizing this transport segment.

Generally, both the models used (NC-FRLM and PyPSA) are greenfield models that generate a perfect foresight outcome without restrictions regarding its pathway. This allows us to abstract from such restrictions and develop an understanding of the optimal future system layout. However, real-life developments towards such ideal states are likely to be not as efficient and therefore should be addressed in future work.

# 7 Conclusions

In this paper, we investigated the interplay of hydrogen refueling stations with the power system on a national level. We combined two methods, a refueling location model and an energy system model, and developed an understanding of how they interact.

Initially, we determined the increase in total installed capacity by applying the refueling location model to the German HDV sector and fuel cell technology. The resulting very limited number of about 140 potential new HRS needed to supply the national fleet results in an additional installed capacity of 60 GW. Using fuel cell HDVs in Germany would increase the total electricity demand by about 70 TWh for national traffic and incur additional infrastructure costs of about 7 billion euro per annum.

Further, we showed that the levelized cost of hydrogen varies regionally depending on the local cost of electricity production and that system-aware dimensioning (i.e. more flexibility) in hydrogen refueling infrastructure reduces the LCOH by 0.77 euro/kg as well as the total system costs by almost 1 billion euro/a in the case study.

We therefore conclude that the co-optimization of multiple energy sectors is important for investment planning in the energy system and promises to exploit synergies and cost reduction potentials if its components act in concert. Moreover, power markets based on nodal pricing pass on vital information about cost-effective energy consumption from a system perspective. This is a prerequisite for investors to consider the system perspective when investing in and sizing, e.g. refueling stations.

## Acknowledgements

Philipp Kluschke gratefully acknowledges funding by the German Federal Ministry of Research and Education (ENavi project, FKZ 03SFK4N0) and the framework of the Profilregion Mobilitätssysteme Karlsruhe, which is funded by the Ministry of Science, Research and the Arts and the Ministry of Economic Affairs, Labour and Housing in Baden-Württemberg and as a national High Performance Center by the Fraunhofer-Gesellschaft.

Fabian Neumann gratefully acknowledges funding from the Helmholtz Association under grant no. VH-NG-1352.

Both authors thank Tom Brown and Martin Wietschel for helpful comments, support and guidance.

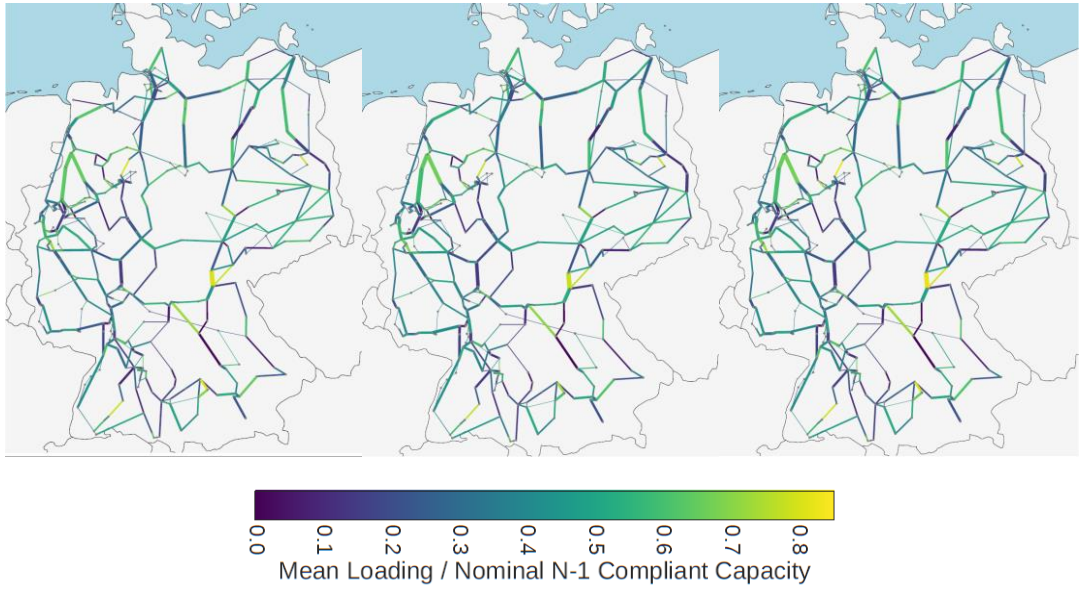

Figure 12: Mean networking loading for scenario without HRS (left), scenario 1 (middle) and scenario 2 (right)

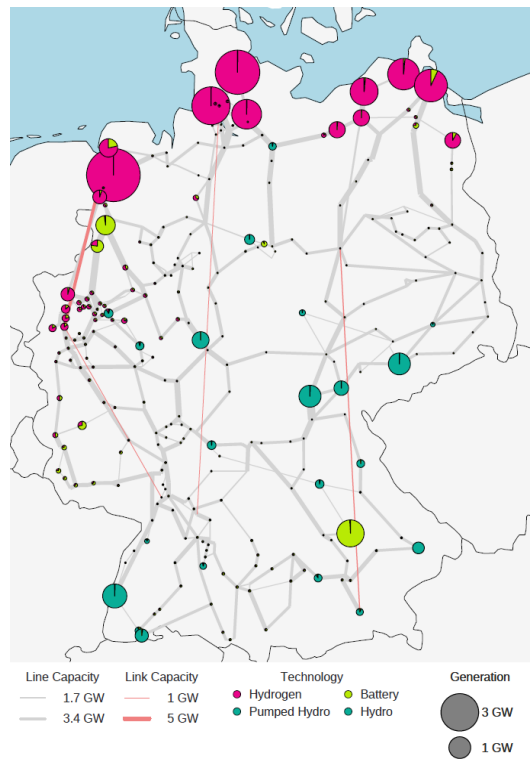

Figure 13a: Network extension and storage power capacities for scenario without HRS

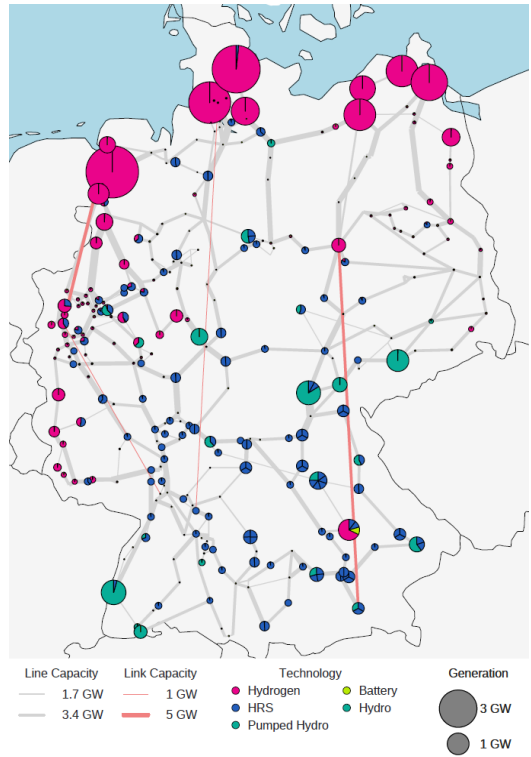

Figure 14b: Network extension and storage power capacities for scenario 1

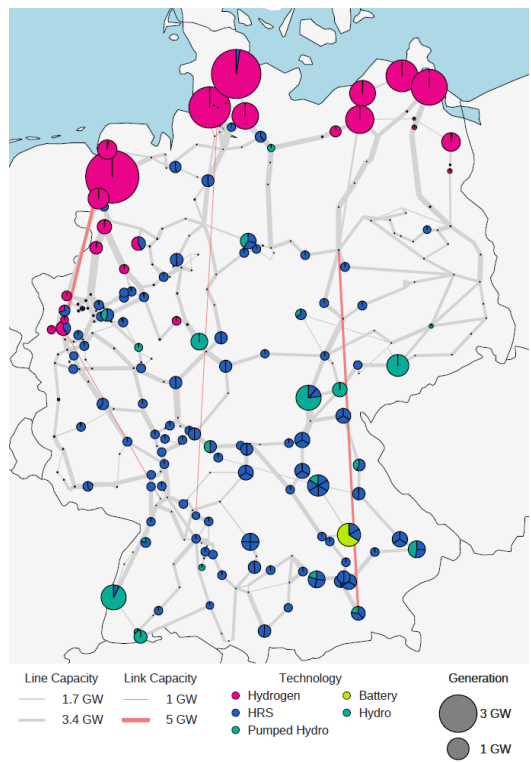

Figure 15c: Network extension and storage power capacities for scenario 2

| Parameter | Unit | Diesel | FCEV | BEV |
|---|---|---|---|---|
| Energy at wheel | [kWh/100_km] | 120.0 | 120.0 | 120.0 |
| Powertrain efficiency | [%] | 0.30 | 0.55 | 0.75 |
| Energy at tank | [kWh/100_km] | 360.7 | 218.2 | 160.0 |
| Energy content | [kWh/l] or [kWh/kg] | 10.0 | 33.3 | - |
| Energy | [l/100_km] or [kg/100_km] | 36.1 | 6.6 | - |
| Fuel price HDV-Diesel (status 02.07.2019) | [euro/l] | 1.2 | - | - |
| Fuel price HDV/Diesel | [euro/100_km] | 43.3 | | |
| *Assumption: Cost parity between Diesel and FCEV & BEV to enable acceptance* | *[euro/100_km]* | | *43.3* | *43.3* |
| *Resulting max. fuel price FCEV* | *[euro/kg_hydrogen]* | - | *6.6* | - |
| *Resulting max. fuel price BEV* | *[euro/kWh_electricity]* | - | - | *0.27* |

Table 5: Analysis of current fuel cost (euro/100km) for HDVs with diesel powertrains and derivation of maximum energy cost for hydrogen (FCEV) and electricity (BEV) to be competitive with current diesel prices

| Parameter | Unit | XS | S | M | L | XL | XXL |
|---|---|---|---|---|---|---|---|
| High pressure storage | [Mio. €] | 0.13 | 0.26 | 0.51 | 1.03 | 2.06 | 2.06 |
| Dispenser | [Mio. €] | 0.11 | 0.11 | 0.21 | 0.43 | 0.86 | 0.86 |
| Compressors | [Mio. €] | 1.58 | 2.76 | 5.52 | 10.65 | 21.30 | 21.30 |
| Colling unit | [Mio. €] | 0.12 | 0.12 | 0.12 | 0.12 | 0.12 | 0.12 |
| Safety Features | [Mio. €] | 0.14 | 0.14 | 0.28 | 0.56 | 1.12 | 1.12 |
| **Total Station Costs** | [Mio. €] | 0.59 | 1.19 | 2.37 | 4.74 | 9.48 | 18.96 |

Table 6: Overview on technology and economics for all HRS types (XS to XXL) without electrolyser and low-pressure storage, based on HDRSAM tool (Elgowainy et al., 2017) and own assumptions for 2050